\documentclass[amsmath]{aastex702}

\usepackage{amssymb,amsfonts}%
\usepackage{amsthm}%
\usepackage{mathrsfs}%
\usepackage{textcomp}%
\usepackage{booktabs}%

\begin{document}

\title{Grid Optimization for Wide Field Modulation Collimators}

\correspondingauthor{Steven E. Boggs}
\author[gname=Steven,sname=Boggs]{Steven E. Boggs}
\affiliation{Department of Astronomy \& Astrophysics, University of California, San Diego, 9500 Gilman Drive, La Jolla, CA 92093, USA}
\affiliation{Space Sciences Laboratory, University of California, Berkeley, 7 Gauss Way, Berkeley, CA 94720, USA}
\email[show]{seboggs@ucsd.edu}

\begin{abstract}
Wide-field localization of transient sources such as gamma-ray bursts (GRBs) places new demands on the grid design of modulation collimators, whose field of view (FOV) is normally limited by self-vignetting. An opaque grid slat thick enough for a given energy also clips off-axis rays, shrinking the effective slit width and setting the FOV. Historical bigrid and rotating-modulation-collimator (RMC) designs, from early wire grids to the photo-etched rectangular slats used for RHESSI, treat this trade-off as geometrically unavoidable. We ask whether the slat's cross-sectional shape can instead be optimized for wide FOV, deriving a transmission-function framework based on an effective slit width that depends on slat geometry, and comparing four cross sections at fixed thickness and pitch: rectangular, an idealized diamond profile, a hexagonal profile approximating photochemical etching, and the circular wire grids of early X-ray astronomy. For the same pitch and thickness, the diamond profile extends the FOV and eliminates off-axis clipping over most of the field; a hexagonal profile captures much of this benefit. Extending the analysis to four-phase demodulation, we show that slat shape sets the extent of resulting phase-space blind spots, and that the diamond profile removes them entirely across a target FOV where the rectangular baseline cannot. Finite grid opacity is a largely separable effect -- leakage decreases off-axis and leaves the blind-spot boundaries unchanged. These results identify slat cross-sectional shape as a design parameter that can substantially increase throughput and widen the blind-spot-free FOV of a wide-field modulation collimator.
\end{abstract}


\section{Introduction}\label{sec1}

Rapid, precise localization of transient hard X-ray and soft gamma-ray
sources such as gamma-ray bursts (GRBs) is a longstanding challenge for
wide-field high-energy astronomy, historically addressed by
coded-aperture mask imagers \citep{Gehrels2004,Kawai2003}, multi-satellite
triangulation \citep{Hurley2013,Burns2023}, and rotation modulation
collimator (RMC) imagers built on the \emph{bigrid} concept introduced by
\citet{Oda1965}, a pair of separated, aligned, periodic slit grids whose
transmitted flux encodes the angular position of an off-axis source
\citep{Lund1986,Lund1995,Hurford2002}. A companion paper
\citep{Boggs2026} reviews this bigrid/RMC heritage and introduces
an alternative architecture built on the same bigrid concept, a fixed
(non-rotating) modulation collimator (FMC) that combines four-phase
demodulation with a vernier cascade of grid angular modulation
frequencies to achieve continuous, wide-field GRB localization with no
moving parts. That paper treats each bigrid pair as an idealized,
perfectly opaque pair of infinitesimally thin slit grids. The present
paper removes that idealization and asks how the grid's own
cross-sectional shape, once it is given the finite thickness opacity
requires, sets the field of view actually achievable.

Oda's original concept, and its earliest flown realizations, used grids
of wound wire. \citet{Gursky1966} located Sco X-1 and the Crab Nebula to
$15''$--$30''$ using sounding-rocket four-grid collimators of this
construction. \citet{Bradt1968} reviewed this modulation-collimator
technique as it stood by the end of that decade. As the technique was
pushed toward finer angular resolution and higher photon energies over
the following years, flat, rectangular-cross-section slats, photo-etched
or electroformed rather than wound or woven, came to replace round wire
grids. By the time of \textit{RHESSI}, whose finest grids reach a
$34\,\mu$m pitch in molybdenum \citep{Hurford2002}, rectangular slat
construction had become the standard for the technique. This shift is a
direct consequence of the limitations of the wire geometry. A wire of diameter $d$ has its
thickness and pitch locked together ($t=p/2=d$; see
Section~\ref{subsec3-4}), so a wire grid cannot be made more opaque to
higher-energy photons without also coarsening its pitch and degrading
angular resolution, whereas a rectangular slat can be made as thick as
opacity requires independently of the pitch set for resolution.

A grid of finite thickness $t$ made of either wires or rectangular slats
necessarily clips off-axis rays once the photon's lateral displacement
across the grid's depth becomes comparable to the slit width $w$, and the
resulting shrinkage of the effective slit width with off-axis angle is
exactly what \citet{SaintHilaire2024} illustrate for RHESSI, whose finest
subcollimators see their effective slit-to-pitch ratio fall substantially
over their $\sim\!1^\circ$ field of view (FOV) as a direct consequence of
shadowing by the slats' finite depth. More generally, this review frames
the resulting relationship between angular resolution, FOV, and the
minimum slat thickness needed for opacity at a given energy as an
essentially unavoidable, purely geometric constraint on a bigrid
collimator, implicitly assuming a rectangular slat cross section.
\citet{Gaither1996} performed a quantitative treatment of the resulting
transmission-function distortion, characterizing the thickness-induced
flattening and new even-harmonic content of a rotating RMC's fringe
pattern. Their treatment, like the rest of the heritage above, takes the
rectangular slat as given. Whether the FOV constraint is a property of
the underlying geometric optics or of the slat shapes conventionally used
to realize it has not, to our knowledge, been examined. The companion
paper introducing the FMC architecture sidesteps
limitation. It treats the grid as infinitesimally thin, explicitly notes
\citeauthor{Gaither1996}'s thickness-induced flattening as an effect
studied in this paper, and adopts a $60^\circ$ target
FOV in its worked example without addressing what grid thickness and slat
shape that FOV actually requires \citep{Boggs2026}. The present paper
addresses these questions directly, and, as we show in
Sec.~\ref{sec4}, adopts the same $60^\circ$ target FOV so that the two
papers' worked examples can be compared directly.

\begin{figure}
\centering
\plotone{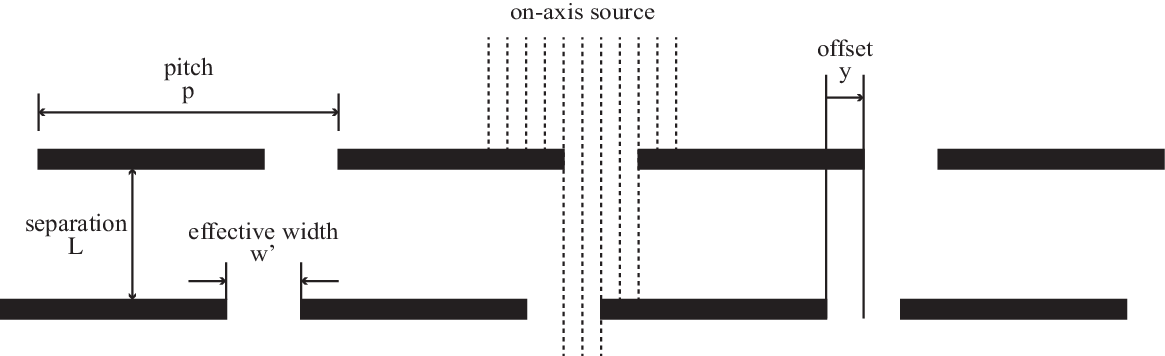}
\caption{\label{fig:f1} Transmission through a bigrid pair can be characterized by the effective width $w'$ of the grid slits, and the relative offset $y$ between the grids. When $y=0$ the slits
are perfectly aligned, allowing the fraction $w'/p$ of the source photons to transmit through the grid. When $y=p$ the grids are perfectly anti-aligned and the transmission is zero.
If $w' < p/2$, the transmission will be zero for a range of $y$ offsets.}
\end{figure}

Outside the bigrid/RMC literature specifically, closely related
self-vignetting problems have been addressed by reshaping or reorienting
the absorber rather than accepting the trade-off. For thick coded-aperture
masks, \citet{GrindlayHong2004} and \citet{Hong2004} proposed tilting each
mask element radially, so that the outermost elements of the mask point towards
the edges of the
wide FOV rather than remaining perpendicular to the mask plane, in order to
suppress the auto-collimation that would otherwise result from the
$\sim\!7\,$mm-thick tungsten mask required for the proposed EXIST mission.
Laboratory tests confirmed a substantial reduction in self-collimation
relative to a conventional perpendicular-hole mask. 
An older and by now
well-established relative of this idea is the ``focused'' anti-scatter
grid used in medical and industrial radiography, in which slats are tilted
to converge on a finite-distance focal spot so that primary rays are not
clipped at the edges of the field (see, e.g., \citealt{Lifton2025} for a
recent industrial CT implementation). Both of these solutions modify the
absorber's \emph{orientation} as a function of position across the
aperture, a spatially varying tilt, sometimes called a focused grid or
mask, rather than the local cross-sectional \emph{shape} of an
individual slat or hole.

In this paper we investigate whether changing the grid slat
\emph{shape}, as opposed to tilting it, can relax the FOV/thickness
trade-off. To our knowledge this question has not been previously
addressed for bigrid or RMC collimators, fixed or rotating. Although our
treatment applies to any bigrid instrument, we adopt the wide-field FMC
of \citet{Boggs2026} as our worked example throughout, since it is
precisely the wide-field case for which the FOV/thickness trade-off is
most consequential. Section~\ref{sec2} develops a general
transmission-function framework, building on the ideal-grid formalism,
in terms of an effective slit width $w'(\theta)$ that
depends on slat geometry; Section~\ref{sec3} evaluates this framework for
rectangular, diamond, hexagonal (approximating a photochemically etched
profile), and circular (wire) slat cross sections; Section~\ref{sec4}
compares their resulting peak transmission across a wide FOV;
Section~\ref{sec5} quantifies the consequences for four-phase
demodulation including potential blind spots; and
Section~\ref{sec6} extends the treatment to partially transparent slats.

\begin{figure}
\centering
\plotone{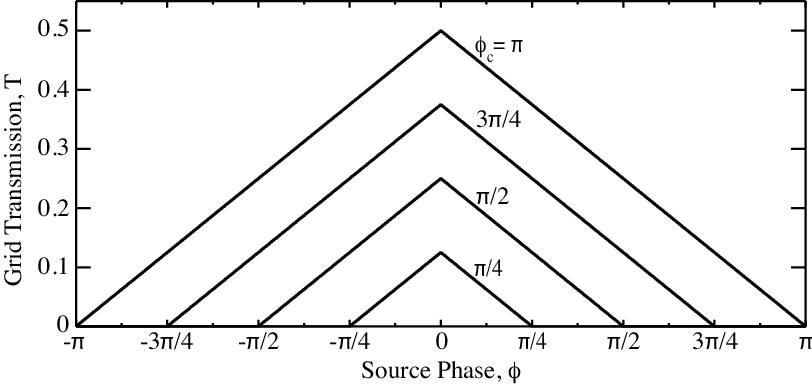}
\caption{\label{fig:trans_w} Transmission function [Eq.\,\ref{eq:trans-phi}]  for varying values of the critical phase $\phi_c$.
The finite thickness of a grid clips photons coming in from off-axis, decreasing $\phi_c$ and flattening the transmission through the grids.}
\end{figure}

\section{Transmission Function}\label{sec2}

Consider the transmission of photons through an idealized perfectly-opaque, infinitely-thin bigrid pair (Fig.\,\ref{fig:f1}). The grid is characterized by a pitch $p$, and an
\emph{effective slit width} $w'$. Here we are introducing this as an ``effective'' width because the concepts introduced in this section can
be carried over to realistic grid designs while generalizing $w'$ as a function of offset source angle $\theta$, i.e., $w'(\theta)$.
This source angle $\theta$ is measured in the one-dimensional offset direction perpendicular to the grid slits.

When the two grids are perfectly
aligned the transmission through the bigrids is maximum, with a peak transmission of $T_{peak} = \frac{w'}{p}$.
This maximum occurs whenever the shadow of the upper slit overlaps perfectly with the lower slit, which occurs whenever
the tangent of the source angle, $\tan \theta$, is an integer multiple of $p/L$.

It is convenient to define the offset from perfect alignment in terms of the phase of the repetition period, the \emph{source phase} $\phi$, which is the quantity actually measured by a single bigrid pair. This is the same source phase measured in \citet{Boggs2026}, where it is defined via the angular fringe offset $\delta$ and modulation period $\alpha$ as $\phi = 2\pi\delta/\alpha$ [their Eqs.~(4)-(5)]. Here we define it instead through a hypothetical physical shift $y$ of the grids relative to each other (Fig.\,\ref{fig:f1}), which is equivalent in the small-angle regime ($\tan\theta\approx\theta$) adopted throughout both papers.
We define $\phi = 0$ when the grid slits are aligned. When the slits of the top grid are centered on the slats of the bottom grid,
the phase is $\phi = \pi$. This offset $y$ is related to the phase $\phi$ through the fact that $\phi$
repeats each time the grids are shifted by an offset $p$. The relation between these two quantities is therefore $\phi = 2 \pi y/p$,
where we generally define $-\pi < \phi \le \pi$. Shifting the grids by a distance $y$ for an on-axis source is equivalent to keeping
the grids aligned and shifting the source to an off-axis angle $\tan \theta = -y/L$.

Using this simple geometry it is straightforward to find the transmission through the grids as a function of phase. In terms of the hypothetical displacement $y$, the transmission is
given by the simple relation:
\begin{equation}
			T(y) = \left\{ \begin{array}{rcl}
			\frac{w'}{p} \left[1-\frac{|y|}{w'}\right] & \mbox{for} & 0 < |y| \le w' \\
			0 & \mbox{for} & w' < |y| < p \mbox{.}
			\end{array}\right.
\label{eq:trans-y}
\end{equation}
\noindent
From this relation we can define a critical phase angle,
\begin{equation}
			\phi_c = \frac{2 \pi w'}{p},
\label{eq:phi_c}
\end{equation}
\noindent
 at which point the transmission drops to zero. In terms of phase, we can write the transmission function,
 \begin{equation}
			T(\phi) = \left\{ \begin{array}{rcl}
			\frac{w'}{p} \left[1-\frac{|\phi|}{\phi_c}\right] & \mbox{for} & 0 < |\phi| \le \phi_c \\
			0 & \mbox{for} & \phi_c < |\phi| < \pi \mbox{.}
			\end{array}\right.
\label{eq:trans-phi}
\end{equation}
\noindent

Figure\,\ref{fig:trans_w} shows this transmission function [Eq.\,\ref{eq:trans-phi}]  for varying values of $\phi_c$.
This function recovers the ideal, triangular (sawtooth) transmission function of \citet{Boggs2026} for traditional (rectangular, infinitesimally thin) modulation collimators when $w' = p/2$. In this limit $\phi_c = \pi$, and the
peak transmission $T_{peak} = \frac{w'}{p} = \frac{1}{2}$.

Whenever $w' < p/2$, the resulting transmission pattern is flattened, leading to phase regions ($\phi_c < |\phi| < \pi$) where the bigrid pair transmission goes to zero.
For a source whose position corresponds to a phase that falls within this region, the bigrid pair is insensitive to the source.
For a single, fixed bigrid pair this would result in blind spots within the instrument FOV. RMC collimators overcome this
shortcoming naturally through their rotation, since the phase of the source will vary with the rotation angle of the instrument.
For our FMC design, this is overcome by combining four-phase demodulation (i.e., four bigrids of equal pitch but varying phases, sampling four different phases
for a given source), and a vernier cascade with bigrids of varying angular modulation frequency.
We will explore these blind spots in more detail in Section\,\ref{sec5}.

\begin{figure}
\centering
\plotone{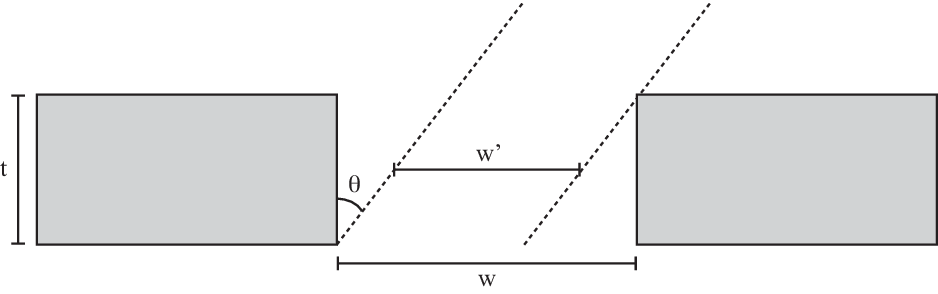}
\caption{\label{fig:f3} The effective slit width $w'$ for rectangular slats. The rays clipped by the slat corners are shown here for $\theta > 0$, which effectively reduces the width of the slit
from $w$ to $w'$.}
\end{figure}

\section{Grid Geometries}\label{sec3}

Building on the transmission model of Section\,\ref{sec2}, we now use the concept of an effective slit width $w'$ to characterize how different slat geometries perform
as a function of source angle $\theta$. Each geometry below defines a different cross-sectional shape for the grid slats. Throughout this section
we treat the grids as perfectly opaque, deferring the effects of finite opacity to Section\,\ref{sec6}. Treating the grids as perfectly opaque here
provides a powerful framework for isolating the effects of slat geometry on performance across a wide FOV.

\subsection{Rectangular Grids}\label{subsec3-1}

Modern modulation collimators have utilized rectangular slats (Fig.\,\ref{fig:f3}). In the standard configuration the grids are $50\%$ open, i.e., $w = p/2$.
We will assume this standard configuration for this derivation.
In this diagram we have shown the left- and right-most rays that will still pass through the slit for a source
at angle $\theta$. Because the grids have a finite thickness $t$, and treating the grid as perfectly opaque,
the corners of the
slats will clip the incoming photons, blocking their passage through the bigrid.
As $\theta$ increases, the effective width $w'$ of the slit
(the horizontal distance between these left- and right-most rays) decreases.
From the rectangular geometry in Figure\,\ref{fig:f3} we can see that
$w' \rightarrow 0$ when $\tan \theta \rightarrow w/t$. This angle defines the FOV $\Omega$ for rectangular grids:
\begin{equation}
\tan \Omega = w/t,
\label{eq:omega-rectangle}
\end{equation}
\noindent
the same self-vignetting angle $\Omega$ defined geometrically, from the grid thickness and slit width alone, as one of the three performance parameters of \citet{Boggs2026} (their Eq.~3), though not propagated there into the transmission function itself. The treatment here recovers that definition exactly for the rectangular case and generalizes it to the other slat geometries of Secs.~\ref{subsec3-2}--\ref{subsec3-4}.

Using this definition for $\Omega$, we can define the effective slit width for rectangular grids:
\begin{equation}
			w'(\theta) = \left\{ \begin{array}{rcl}
			w - t/\tan \theta & \mbox{for} & 0 < \theta \le \Omega \\
			0 & \mbox{for} & \Omega < \theta \mbox{.}
			\end{array}\right.
\label{eq:w-rectangle}
\end{equation}
\noindent

Now that we have defined $w'(\theta)$ for the rectangular grids, we can plug this function back into our transmission function [Eq.\,\ref{eq:trans-phi}]
to solve for the transmission function for these rectangular grids across the FOV of the instrument. We explore this peak transmission function for all
the grid geometries in Section\,\ref{sec4}.

\begin{figure}
\centering
\plotone{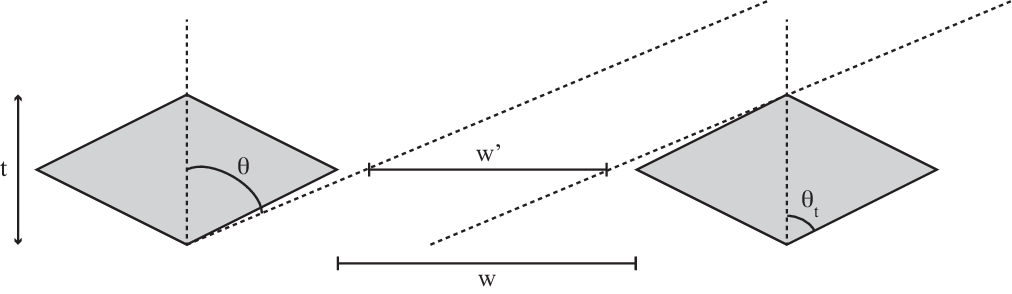}
\caption{\label{fig:f4} The effective slit width for diamond slats. The clipped rays are shown here for $\theta > \theta_t$, reducing the effective width of the slit.
For $\theta \le \theta_t$, the effective width does not depend on the source angle, and is just $w' = w$, the physical width of the slit.}
\end{figure}

\subsection{Diamond Grids}\label{subsec3-2}

For the rectangular geometry, both the limited FOV and the decrease in $w'$ at off-axis angles are due to the slat corners
clipping incoming photons, an effect with an especially large impact on grids designed for wide-field instruments.
The natural solution to this clipping (setting aside, for now, whether such a geometry could be manufactured) is to remove the corners entirely,
creating a diamond cross-section (Fig.\,\ref{fig:f4}).

The key feature of this geometry is that for angles $\tan \theta \le \frac{(w/2)}{(t/2)}$ the incoming photons are not clipped by any corners,
so $w' = w$. We define this transition angle,
\begin{equation}
\tan \theta_t = \frac{w}{t}.
\label{eq:w-diamond-transition}
\end{equation}
\noindent
Note that this transition angle is the same as the FOV for rectangular grids,
but with the distinction that this angle defines where the grid clipping \emph{begins} for diamond grids, as opposed to where the FOV \emph{ends} for
rectangular grids.

For angles $\theta > \theta_t$, the incoming photons get clipped by the top and bottom points of the diamonds (Fig.\,\ref{fig:f4}). This is similar
to the clipping for the rectangular grid, but between points separated by $2w$. This gives an effective width in this range $w' = 2w - t\,\tan \theta$.
This clipping leads to $w' = 0$ at the angle $\tan \theta = 2w/t$, which defines the FOV of the diamond grids:
\begin{equation}
\tan \Omega = 2w/t.
\label{eq:omega-diamond}
\end{equation}
\noindent

Combining these regions together, the full equation for the effective width $w'$ for diamond grids is given by:
\begin{equation}
			w'(\theta) = \left\{ \begin{array}{rcl}
			w & \mbox{for} & 0 < \theta \le \theta_t \\
			2w - t\,\tan \theta & \mbox{for} & \theta_t < \theta \le \Omega \\
			0 & \mbox{for} & \Omega < \theta \mbox{,}
			\end{array}\right.
\label{eq:w-diamond}
\end{equation}
\noindent
where we note that $\Omega$ is larger for the diamond grids compared to the rectangular grids.

Already this reveals a strong advantage of diamond grids over rectangular grids for wide-field instruments. For rectangular grids, the
transmission function starts to flatten immediately for any off-axis source.
By contrast, diamond grids maintain an unflattened transmission function for all source angles less than $\theta_t$, and the instrument FOV,
defined as where the effective slit width $w'$ goes to zero, is nearly
double that of the rectangular grids of equal pitch and thickness.

\subsection{Hexagonal Grids}\label{subsec3-3}

\begin{figure}
\centering
\plotone{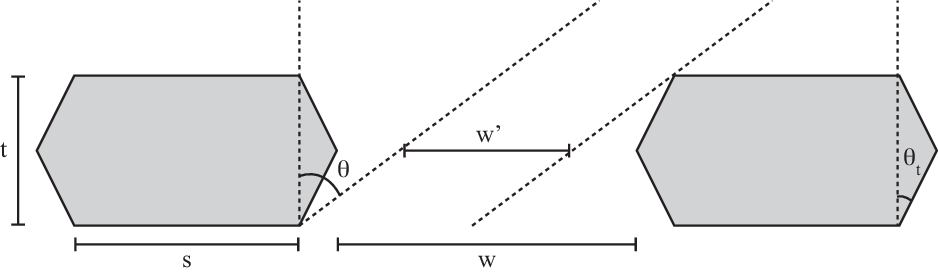}
\caption{\label{fig:f5} The effective slit width for hexagon slats. The clipped rays are shown here for $\theta > \theta_t$, reducing the effective width of the slit.
For $\theta \le \theta_t$, the effective width does not depend on the source angle, and is just $w' = w$, the physical width of the slit, similar to diamond slats.}
\end{figure}

The diamond slats discussed above are an ideal shape for wide-field modulation collimators. However, a more realistic shape in terms of manufacturability of the
grids is an irregular hexagonal shape (Fig.\,\ref{fig:f5}), where the rectangular corners are removed but the slat remains partially rectangular in profile. These slats are
characterized by a width $s < w$ at the top and bottom of the slat, and a width through the middle horizontal plane of $w = p/2$.

For this profile we can define a transition angle $\theta_t$,
\begin{equation}
 \tan \theta_t = \frac{(w-s)/2}{t/2} = \frac{w-s}{t}.
\label{eq:theta-t-hex}
\end{equation}
\noindent
This geometry is similar to the diamond profile for angles $\theta \le \theta_t$ because the incoming photons do not get clipped by any corners,
so $w' = w$.

For angles $\theta > \theta_t$, the upper and lower corners again clip the photons, similar to the rectangular case, but now with the corners separated by $2w-s$.
This gives an effective width in this range $w' = (2w - s) - t\,\tan \theta$. This effective width goes to zero when $\tan \theta = (2w-s)/t$, defining the FOV for these hexagonal grids:
\begin{equation}
\tan \Omega = \frac{2w-s}{t}.
\label{eq:omega-hexagon}
\end{equation}
\noindent

Combining these regions together, the full equation for the effective width $w'$ for hexagon grids is given by:
\begin{equation}
			w'(\theta) = \left\{ \begin{array}{rcl}
			w & \mbox{for} & 0 < \theta \le \theta_t \\
			(2w - s) - t\,\tan \theta & \mbox{for} & \theta_t < \theta \le \Omega \\
			0 & \mbox{for} & \Omega < \theta \mbox{,}
			\end{array}\right.
\label{eq:w-hexagon}
\end{equation}
\noindent
where again $\Omega$ is uniquely defined above for this geometry.

As a consistency check, both $w'$ and $\Omega$ agree with the rectangular solutions in the limit $s \rightarrow w$, and with the diamond solutions in the limit $s \rightarrow 0$.

The hexagon grid shows some of the advantages of the diamond grids over rectangular grids, such as maintaining an unflattened transmission function for source angles $\theta \le \theta_t$,
but not over the entire equivalent FOV of a rectangular grid like the diamond grids do. The hexagon grids also extend the FOV beyond the limits of the rectangular grid
of equal pitch and thickness.

\subsection{Circular Grids}\label{subsec3-4}

\begin{figure}
\centering
\plotone{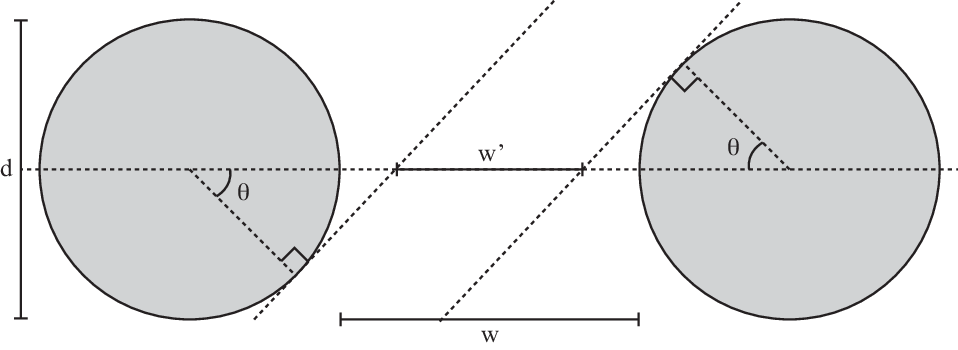}
\caption{\label{fig:f6} The effective slit width $w'$ for circular slats (wires). Rays are clipped when tangential to the circle, reducing the effective width of the slit.
Similar to the rectangular grid, rays are clipped ($w' < w$) for all angles $\theta > 0$}
\end{figure}

The original grids used in X-ray
astronomy were made of round wires strung in a repeating pattern, so with circular cross-sectional area (Fig.\,\ref{fig:f6}).
We can solve this case for comparison with our other geometries. We will assume that the wires have a diameter $d$, and
are equally-spaced with a pitch $2d$. So in comparison with our previous geometries, $t=d$ and $w=d$.

For this geometry, there are no corners to clip incoming rays, but rays are clipped where they become tangent to
the circular wires (Fig.\,\ref{fig:f6}). This gives an effective width in this range $w' = 2w - w/\cos \theta$.
This clipping leads to $w' = 0$ at the angle $\cos \theta = 1/2$, defining the FOV for these circular grids:
\begin{equation}
\Omega = \frac{\pi}{3}.
\label{eq:omega-circle}
\end{equation}
\noindent

Combining these regions together, the full equation for the effective width $w'$ for circular grids is given by:
\begin{equation}
			w'(\theta) = \left\{ \begin{array}{rcl}
			w(2-1/\cos \theta) & \mbox{for} & 0 < \theta \le \pi/3 \\
			0 & \mbox{for} & \pi/3 < \theta \mbox{,}
			\end{array}\right.
\label{eq:w-circle}
\end{equation}
\noindent
where $w=d$, the wire spacing is equal to the wire diameter.

The transmission function for circular grids starts to flatten immediately for off-axis sources, but not as quickly as the rectangular grids due to not having the sharp corners.
While the manufacture of circular grids out of wires is relatively straightforward, they suffer from two major disadvantages. First, the diameter of the wires set both the thickness
and the pitch of the grid, leaving little room for tuning the grid response in terms of collimated energy and angular resolution.
Also, the circular wire grids have a fixed FOV, regardless of the diameter of the wires, again limiting the ability
to tune the grid response.

\section{Peak Transmission}\label{sec4}

Having defined the effective slit width $w'(\theta)$ for each of the grid geometries, we can return to the transmission function in Section\,\ref{sec2}. From Eq.\,\ref{eq:trans-phi},
we can define the peak transmission, $T_{peak}(\theta)$, which is the transmission when the slits are perfectly aligned, i.e., the source phase $\phi = 0$:
\begin{equation}
 T_{peak}(\theta) = \frac{w'(\theta)}{p},
\label{eq:t-peak}
\end{equation}
\noindent
where $w'(\theta)$ is unique for each of the grid geometries discussed above.

\begin{figure}
\centering
\plotone{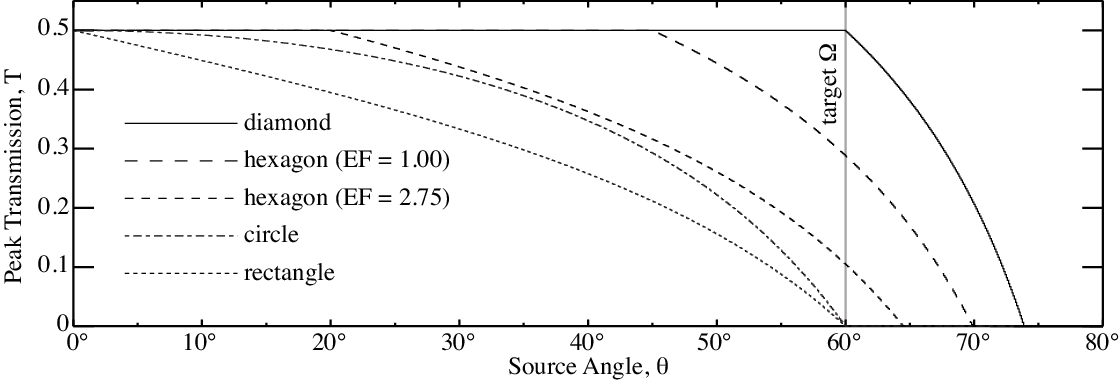}
\caption{\label{fig:peak_trans} Peak transmission ($w'/p$) as a function of the source off-axis angle $\theta$. Shown for comparison is the target FOV ($\Omega = 60^\circ$), defined
by the rectangular grid for comparison to the other geometries utilizing the same grid pitch $p$ and thickness $t$. The diamond grid shows superior performance at all source angles.}
\end{figure}

These peak transmission curves are plotted in Figure\,\ref{fig:peak_trans}. For this work we keep the ratio $w/t$ equal for each
grid design, and set a target FOV $\Omega = 60^\circ$, the same field of view adopted in the worked example of \citet{Boggs2026} (their Sec.~6.4).
These curves were generated assuming $w/t = \tan 60^\circ$, corresponding
to a FOV $\Omega = 60^\circ$ for the rectangular grid. Every other geometry in this section shares this same $w$ and $t$, so any difference
in FOV or transmission among them is due entirely to slat shape.

The plot draws a stark contrast between the rectangular grid and the diamond grid. For the same grid thickness $t$ and slit width $w$, the rectangular grid's peak transmission drops
immediately for $\theta > 0$, reaching zero at $\theta = 60^\circ$, precisely the FOV that \citet{Boggs2026} adopts as a target without deriving what grid geometry would be required to reach it. By contrast, the diamond grid exhibits constant peak transmission over the entire
region $\theta \le 60^\circ$, and only then begins dropping, eventually reaching zero at $\theta \simeq 74^\circ$.

The hexagon grids require an additional parameter, the slope of the slits $\theta_t$, which also defines the top and bottom slat width $s$ (Fig.\,\ref{fig:f5}).
This irregular hexagon approximates the shape that can be achieved through photochemical etching. This process cannot create
perfectly rectangular grids, which is actually an advantage for wide-field designs. The slope of these etched faces is defined by the etching factor ($EF$) of the etching process.
This etching factor is defined by the relation:
\begin{equation}
 EF \simeq \frac{t}{w-s},
\label{eq:EF}
\end{equation}
\noindent
which is related to our transition angle by the simple relation $\tan \theta_t = 1/EF$. This relation for $EF$ is approximate because photochemical etching does not result in perfectly straight edges.
However, the irregular hexagon model is a close approximation.

Typical manufacturers quote etching factors ranging from $EF \sim 1.00$ to $EF \sim 2.75$, both of which are plotted in Figure\,\ref{fig:peak_trans}. A larger $EF$ corresponds to steeper walls on the etch, i.e.,
a nearly-perfect rectangular grid would require $EF \gg 1$. Values of $EF \sim 1$ are optimal for replicating many of the advantageous
features of the diamond grids for wide-field applications.

For comparison, the circular (wire) grids do show better off-axis performance overall than the rectangular grids,
and could remain a viable option for wide-field applications that do not require tuning. Notably, the fixed FOV of a circular grid,
$\Omega = \pi/3 = 60^\circ$ [Eq.~\ref{eq:omega-circle}], coincides exactly with the target FOV adopted throughout this section.
Unlike the diamond or $EF\sim1$ hexagon profiles, however, a circular grid only reaches that FOV with its
peak transmission already substantially degraded from the on-axis value, as Figure\,\ref{fig:peak_trans} shows.

\section{Blind Spots}\label{sec5}

\begin{figure}
\centering
\includegraphics[width=0.65\linewidth]{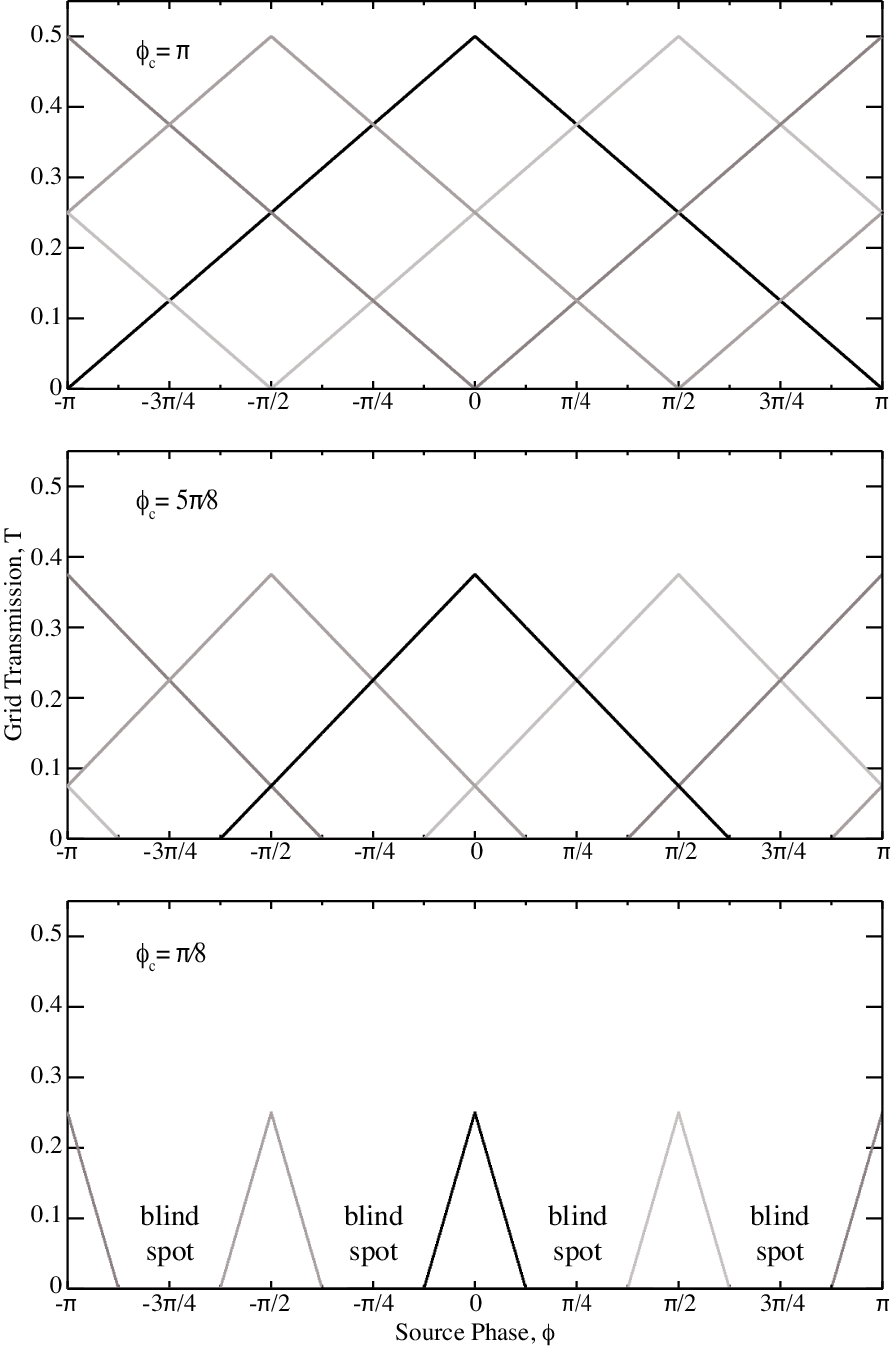}
\caption{\label{fig:blind_spots} Grid transmission for a four-phase module, where the four bigrids are offset in phase by $\pi/2$ relative to each other. (Top) When the response is not
flattened ($\phi_c = \pi$) all four bigrid phases measure modulated source counts for every value of source phase $\phi$. (Middle) For intermediate values of $\phi_c$, only a subset of
the bigrids will measure modulated counts for a given source
phase, while the other bigrids will be ``blind'' to that particular source position.
(Bottom) When $\phi_c < \pi/4$, there will be regions of the source phase where none of the bigrids will measure modulated source counts, leading to blind spots in the FOV.}
\end{figure}

In the reference FMC design of \citet{Boggs2026}, each angular scale is sampled by four bigrid/detector pairs, a four-phase module,
where each of the bigrids is offset in phase by $90^\circ$.
For an ideal measurement, source flux is measured (i.e., modulated) in all four of these bigrid phases, allowing the cleanest measurement of the source phase.
The flattening of the transmission function for off-axis sources (Fig.\,\ref{fig:trans_w}) means that for each bigrid phase there are regions of the source phase
space where that grid response is zero, so that a usable source signal is not measured for that bigrid phase.
This is demonstrated in Figure\,\ref{fig:blind_spots}, where we now plot the transmission function for each of the four phases of the four-phase module.

When the transmission function is not clipped (i.e., $\phi_c = \pi$), there is no flattening of the response, so all four bigrids measure modulated source
counts, regardless of source phase $\phi$ (Fig.\,\ref{fig:blind_spots}, top).
When $3\pi/4 \le \phi_c < \pi$, one of the four bigrids has the potential to be non-responsive to a source.
When $\pi/2 \le \phi_c < 3\pi/4$, two of the four bigrids have the potential to be non-responsive to a source (Fig.\,\ref{fig:blind_spots}, middle).
When $\phi_c < \pi/4$, none of the four bigrid phases can be
guaranteed to be active, creating blind spots within the FOV (Fig.\,\ref{fig:blind_spots}, bottom).
Note, however, that the measurement from a single blind bigrid phase (i.e., where the source falls in the flattened region)
is still useful for measuring the background and source leakage counts that will need to be subtracted from the other
phases. And the null response indicates that the source phase is not consistent with the still-active phase region of that bigrid.

\begin{figure}
\centering
\plotone{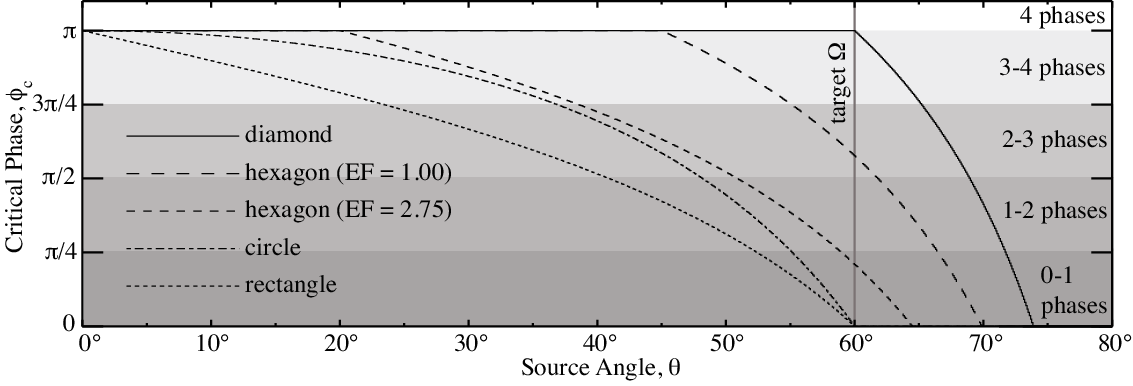}
\caption{\label{fig:crit_phase} Critical phase ($\phi_c$) as a function of the source off-axis angle $\theta$ for the different grid designs. The shaded regions
indicate the number of bigrid phases actively modulating sources for the four-phase module.}
\end{figure}

Optimally, a wide-field grid design will maximize regions where all four bigrid phases are
modulated, and avoid regions where all four bigrids can be potentially blind to a source.
In Figure\,\ref{fig:crit_phase} the number of actively-modulating grids is demonstrated across the FOV of a four-phase module.
This plot makes clear that a major advantage of the diamond grid, beyond maximizing the peak transmission across a wide FOV, is also to extend
four-phase demodulation across the widest field. For this design with a target FOV of $60^\circ$,
the diamond grid design maintains four-phase modulation over the entire target FOV.
This is a sharp contrast to the rectangle grid, which reaches blind spots for significant regions of the
wide-field response. The hexagon grid can significantly improve performance over the rectangular grid, potentially extending four-phase demodulation to off-axis sources
out to $45^\circ$ ($EF = 1.00$), and avoiding blind spots entirely in the target FOV.

\section{Partially Transparent Grids}\label{sec6}

We can characterize the effects of leakage through the grids using an average grid thickness $t_{ave}$ for each of the grid designs. Focusing on
the hexagon grid, the thickness of the slat averaged over the full width $w$ is given by,
\begin{equation}
 t_{ave} = t\frac{(w+s)}{2w},
\label{eq:t_ave}
\end{equation}
\noindent
where $t$ is thickness and $s$ is the slat top and bottom widths (Fig.\,\ref{fig:f5}). This equation
also applies to rectangular grids ($s=w$) and diamond grids ($s=0$).

In the companion paper we established that leakage through the grid is characterized by the factor $\mu x$, where $\mu$ is the linear attenuation
coefficient for the specific photon energy of interest, and $x$ is an average path length of an incident photon through a single grid \citep{Boggs2026}.
We can now define $x$. For a photon originating from a source at angle $\theta$, the average path length through the grid is
given by,
\begin{equation}
x \simeq t_{ave}/\cos \theta.
\label{eq:x_ave}
\end{equation}
\noindent
This approximation holds for grids where $w > t$ and source angles $\theta < 90^\circ$, which apply
to all of the grid designs being considered in this work (rectangular, diamond, hexagon).

Given $\mu x$, we showed in the companion paper that we can write this transmission function using a sawtooth function $h(\phi)$,
\begin{equation}
T(\phi) = \frac{a}{2}\!\left[h(\phi) + 1\right] + \ell,
\label{eq:T-real}
\end{equation}%
where $h(\phi)$ is normalized as ($-1 \le h(\phi) \le 1$), $a \le 1/2$ is the reduced modulation amplitude due to partial grid transparency, and $\ell$ is a leakage term:
\begin{align}
a &= \frac{1}{2} \left( 1-e^{-\mu x}\right)^2,
  \label{eq:leak-a}\\[4pt]
\ell &= e^{-\mu x}.
  \label{eq:leak-ell}
\end{align}%

The primary implication of these equations is that in terms of \emph{leakage} specifically, the grid performance \emph{improves} at off-axis angles, with $a$ increasing
towards $1/2$ and $\ell$ decreasing towards zero as $\theta$ increases. This separates the grid's geometric clipping effects from the effects of finite photon attenuation.
As long as the grid opacity and thickness is being optimized for on-axis sources, the off-axis transparency only decreases, which we have treated as a separate question
from whether the grid geometry itself is optimized for off-axis angles.

While partial transparency affects the overall amplitude of the modulation for a grid, it does not flatten the transmission function like the clipping due to finite thickness grids.
The partial transparency does not affect the critical phase $\phi_c$ for the grid.
Hence, the regions permitting four-phase modulation (Fig.\,\ref{fig:crit_phase}) are not affected by partial transparency of the grids.

\section{Discussion}\label{sec7}

Our results demonstrate that a bigrid collimator's field of view is not
a fixed consequence
of the geometric optics of a thick, opaque grid. For wide-field applications,
the FOV limitations imposed by rectangular slats can be overcome using diamond (ideally) or hexagon (realistically) grid designs.
Holding the pitch $p$, thickness
$t$, and slit width $w$ fixed, and therefore holding the angular resolution
and the on-axis transmission $T_{peak}(0) = w/p$ fixed,
reshaping the slat alone is enough to remove the off-axis
clipping that a rectangular slat introduces immediately at $\theta > 0$.
The corresponding FOV increases
from $\tan\Omega = w/t$ for a rectangular slat to $\tan\Omega = 2w/t$ for
a diamond slat, a factor-of-two increase in $\tan\Omega$ that, for the
$w/t = \tan 60^\circ$ example used throughout Secs.~\ref{sec4}--\ref{sec6},
corresponds to widening the angular FOV from $60^\circ$ to $\sim\!74^\circ$
(Fig.~\ref{fig:peak_trans}). The more consequential change is not this
increase of $\Omega$ itself, but the elimination of flattening
below $\theta_t = \Omega_{\rm rect}$. A diamond-slat grid delivers its full
on-axis peak transmission everywhere within the corresponding entire FOV
of a rectangular grid, rather than degrading monotonically across it.

An idealized diamond slat is unlikely to be achievable with standard
photochemical etching processes. The hexagonal profile
of Sec.~\ref{subsec3-3} is the practical response to this fabrication challenge.
The hexagonal slat is, in
effect, the family of shapes that photochemical etching naturally
produces once a nominally rectangular pattern is over-etched by a
controlled amount, parameterized by the vendor-quoted etching factor
$EF$. A realistic $EF\sim1$ hexagon recovers most of the diamond
slat's benefit, and is sufficient to eliminate the
worst failure mode identified in Sec.~\ref{sec5}, the appearance of
regions where \emph{all four} phases of a four-phase demodulation module
go blind.
Sec.~\ref{sec5}'s finding that blind spots are avoided at $EF\sim1$ over the full
$60^\circ$ target FOV, even though the hexagon's peak-transmission curve
continues to decline beyond $\theta_t = 45^\circ$, illustrates that the
transmission curve alone (Fig.~\ref{fig:peak_trans}) understates the
practical benefit. What determines usable FOV for a four-phase design is
whether at least one of the four phases remains live, a substantially
weaker requirement than maintaining full peak transmission at every
angle.

The circular (wire) grid occupies an instructive middle ground. Its
rounded profile avoids sharp-corner clipping and
consequently outperforms the
rectangular grid in peak transmission at every off-axis angle
(Fig.~\ref{fig:peak_trans}), though, as noted in Sec.~\ref{sec4}, only by
reaching its fixed $60^\circ$ FOV with substantially reduced on-axis
transmission. But because a wire's thickness, width, and
pitch are all set by its diameter $d$, its FOV is
pinned to $\Omega = \pi/3$ regardless of design choice
(Eq.~\ref{eq:omega-circle}), i.e.,
energy response and angular resolution
cannot be tuned independently.
This is precisely the coupling that motivated the historical move
away from wire grids described in Sec.~\ref{sec1}, and it is worth noting
that the diamond and hexagonal profiles solve both problems at once.
They decouple slat thickness (opacity, and hence energy response) from
pitch (angular resolution) in the same way a rectangular slat does, while
additionally decoupling both of these from off-axis FOV performance, a
third design axis that rectangular slats leave unaddressed.

Sec.~\ref{sec6}'s treatment of finite grid opacity shows that this third
performance option can be optimized independently of the first two. Because the
average photon path length through a slat, $x = t_{ave}/\cos\theta$,
increases with off-axis angle, both the leakage $\ell = e^{-\mu x}$ and
the reduced modulation amplitude $a$ improve monotonically off-axis for
\emph{any} of the four geometries. This is the opposite trend from the
geometric clipping effect analyzed in Secs.~\ref{sec3}--\ref{sec5}. Since
partial transparency changes only the overall scale of the transmission
function and not its shape, it leaves the critical phase $\phi_c$, and
therefore the blind-spot structure of Fig.~\ref{fig:crit_phase}, entirely
unchanged. In practice this means the slat shape and etching factor can
be chosen purely to satisfy the FOV and blind-spot requirements of
Secs.~\ref{sec3}--\ref{sec5}, while the material thickness and grid pitch can then be
optimized independently to satisfy the instrument's target energy and angular resolution.

Several simplifications in this treatment are worth noting as
limitations. We have
treated the grid edges as geometrically sharp and perfectly aligned,
neglecting diffraction (relevant at the lowest energies for the finest
pitches; see, e.g., the treatment in \citealt{SaintHilaire2024}) and
fabrication or alignment tolerances beyond the idealized cross-sectional
shapes of Sec.~\ref{sec3}. 
These are real practical concerns, but not
open questions. Grids with sub-$50\,\mu$m photo-etched,
precision-stacked foil pitches and well-characterized etching factors
have already been fabricated, aligned, and flown on \textit{RHESSI} and
\textit{STIX} \citep{Lin2002,Krucker2020}, and the medical-imaging
community has independently developed lithographic and electroforming
techniques capable of micron-level sidewall precision for both parallel
and focused septa geometries \citep{Boone2002,Makarova2002,Makarova2008,VanAudenhaege2015}.
The manufacturing and alignment tolerances needed to realize the
shaped-slat benefits derived here therefore appear to fall within
demonstrated capability, though a dedicated tolerance analysis is
beyond the geometric scope of this paper.
Additionally, the opacity analysis of
Sec.~\ref{sec6} is carried out at a single representative attenuation
coefficient $\mu$. A real instrument operates over a bandpass across
which $\mu$ varies substantially, so while the qualitative direction of
the off-axis leakage improvement is energy-independent, its magnitude is
not, and a full mission design would need to evaluate it across the band
of interest. Neither of these simplifications change the
central geometric result, that slat shape has major effects on the
performance of wide FOV modulation collimators.

\section{Conclusion}\label{sec8}

We have shown that the field-of-view limitation associated with the
finite thickness of bigrid and RMC grid slats is not solely a consequence
of geometric optics, but depends strongly on the cross-sectional shape of
the grid slats, a design parameter left largely unexamined throughout the heritage of
the technique. Using a general transmission-function framework built
around an effective slit width $w'(\theta)$ (Sec.~\ref{sec2}), we derived
closed-form expressions for four slat geometries -- rectangular, diamond,
hexagonal, and circular (wire) -- for a fixed grid pitch and thickness
(Sec.~\ref{sec3}). An idealized diamond profile
eliminates off-axis clipping entirely across the corresponding full FOV of a rectangular
grid, and extends $\tan\Omega$ by a factor of two beyond it
(Sec.~\ref{sec4}). A hexagonal profile with an etching factor of
$EF\sim1$, readily achievable with standard photochemical etching,
recovers most of this benefit and is sufficient to eliminate blind spots
entirely across a $60^\circ$ target FOV in a four-phase demodulation
scheme, where a conventional rectangular grid of the same pitch and
thickness fails over a substantial fraction of that same FOV
(Sec.~\ref{sec5}). Finally, finite grid opacity acts as a separable
effect that improves, rather than degrades, off-axis performance, and
leaves the blind-spot structure unchanged, so the geometric optimization
developed here can be carried out independently of the material and
thickness chosen to satisfy on-axis opacity requirements at a given
energy (Sec.~\ref{sec6}).

Taken together, these results identify slat cross-sectional shape as an
essential design parameter for wide-field modulation
collimators. A grid design that departs from the conventional rectangular
slat can substantially widen the usable, blind-spot-free field of view.
This is directly applicable to the fixed modulation collimator
architecture introduced in our companion work \citep{Boggs2026}, and more
generally to any wide-field hard X-ray or gamma-ray instrument built on
bigrid or RMC principles.

\begin{acknowledgments}
During the preparation of this manuscript the author used Claude (Anthropic, Claude Sonnet 5) to assist with editing text and checking internal consistency.
The ultimate responsibility for the scientific content and integrity of this work rests solely with the human author.
\end{acknowledgments}

\bibliography{refs}{}

@ARTICLE{Boggs2026,
       author = {{Boggs}, Steven E.},
        title = "{Wide Field Localization Using Fixed Modulation Collimators}",
      journal = {arXiv e-prints},
         year = 2026,
        month = sep,
          eid = {arXiv:2609.10384},
        pages = {arXiv:2609.10384},
          doi = {10.48550/arXiv.2609.10384},
archivePrefix = {arXiv},
       eprint = {2609.10384},
 primaryClass = {astro-ph.IM},
       adsurl = {https://ui.adsabs.harvard.edu/abs/2026arXiv260910384B}
}

@incollection{SaintHilaire2024,
  author    = "Saint-Hilaire, Pascal and Shih, Albert Y. and Hurford, Gordon J. and Dennis, Brian",
  title     = "Grid-Based Imaging of X-rays and Gamma Rays with High Angular Resolution",
  booktitle = "Handbook of X-ray and Gamma-ray Astrophysics",
  editor    = "Bambi, C. and Santangelo, A.",
  publisher = "Springer Nature",
  address   = "Singapore",
  pages     = "1783--1816",
  year      = "2024",
  doi       = "10.1007/978-981-19-6960-7_170"
}

@article{Gaither1996,
  author  = "Gaither, C. C., III and Schmahl, E. J. and Crannell, C. J. and Dennis, B. R. and Lang, F. L. and Orwig, L. E. and Hartman, C. N. and Hurford, G. J.",
  title   = "Quantitative Characterization of the X-Ray Imaging Capability of Rotating Modulation Collimators with Laser Light",
  journal = "Appl. {O}pt.",
  volume  = "35",
  number  = "34",
  pages   = "6714--6720",
  year    = "1996"
}

@inproceedings{GrindlayHong2004,
  author    = "Grindlay, J. and Hong, J.",
  title     = "Optimizing Wide-Field Coded Aperture Imaging: Radial Mask Holes and Scanning",
  booktitle = "Optics for EUV, X-Ray, and Gamma-Ray Astronomy",
  editor    = "Citterio, O. and O'Dell, S. L.",
  series    = "Proc. SPIE",
  volume    = "5168",
  pages     = "402--410",
  year      = "2004",
  doi       = "10.1117/12.506260",
  eprint    = "astro-ph/0310474",
  archivePrefix = "arXiv"
}

@INPROCEEDINGS{Hong2004,
       author = {{Hong}, Jae Sub and {Vadawale}, Santosh V. and {Zhang}, Minhua and {Bellm}, Eric C. and {Yousef}, Andrew and {Noss}, John and {Grindlay}, Jonathan E. and {Narita}, Tomohiko},
        title = "{Laboratory coded-aperture imaging experiments: radial hole coded masks and depth-sensitive CZT detectors}",
    booktitle = {Hard X-Ray and Gamma-Ray Detector Physics VI},
         year = 2004,
       editor = {{Burger}, Arnold and {James}, Ralph B. and {Franks}, Larry A.},
       series = {Society of Photo-Optical Instrumentation Engineers (SPIE) Conference Series},
       volume = {5540},
        month = oct,
        pages = {63-72},
          doi = {10.1117/12.559650},
archivePrefix = {arXiv},
       eprint = {astro-ph/0409002},
 primaryClass = {astro-ph},
       adsurl = {https://ui.adsabs.harvard.edu/abs/2004SPIE.5540...63H}
}

@article{Lifton2025,
  author  = "Lifton, Joseph John and Tan, Zheng Jie and Filemon, Christian",
  title   = "Performance of a Focused 2{D} Anti-Scatter Grid for Industrial X-Ray Computed Tomography",
  journal = "J. X-Ray Sci. Technol.",
  volume  = "33",
  number  = "3",
  pages   = "651--661",
  year    = "2025",
  doi     = "10.1177/08953996251325072"
}

@article{Oda1965,
  author  = "Oda, M.",
  title   = "High-Resolution X-Ray Collimator with Broad Field of View for Astronomical Use",
  journal = "Appl. {O}pt.",
  volume  = "4",
  number  = "1",
  pages   = "143",
  year    = "1965"
}

@article{Gursky1966,
  author  = "Gursky, H. and Giacconi, R. and Gorenstein, P. and Waters, J. R. and Oda, M. and Bradt, H. and Garmire, G. and Sreekantan, B. V.",
  title   = "A Measurement of the Location of the X-Ray Source Sco X-1",
  journal = "Astrophys. J.",
  volume  = "146",
  pages   = "310",
  year    = "1966",
  note    = "Reprinted in {Proc.} 1969 conference volume, {ADS} bibcode 1969qhea.conf..215G; the 1966 {ApJ} publication is the primary/citable source"
}

@article{Bradt1968,
  author  = "Bradt, H. and Garmire, G. and Oda, M. and Spada, G. and Sreekantan, B. V. and Gorenstein, P. and Gursky, H.",
  title   = "The Modulation Collimator in X-Ray Astronomy",
  journal = "Space Sci. Rev.",
  volume  = "8",
  pages   = "471--506",
  year    = "1968"
}

@article{Lund1995,
  author  = "Lund, N. and Brandt, S.",
  title   = "Wide Field Monitoring of the X-Ray Sky Using Rotation Modulation Collimators",
  journal = "Exp. Astron.",
  volume  = "6",
  pages   = "19--24",
  year    = "1995"
}

@article{Hurford2002,
  author  = "Hurford, G. J. and Schmahl, E. J. and Schwartz, R. A. and Conway, A. J. and Aschwanden, M. J. and Csillaghy, A. and Dennis, B. R. and Johns-Krull, C. and Krucker, S. and Lin, R. P. and McTiernan, J. and Metcalf, T. R. and Sato, J. and Smith, D. M.",
  title   = "The {RHESSI} Imaging Concept",
  journal = "Sol. Phys.",
  volume  = "210",
  pages   = "61--86",
  year    = "2002"
}

@INPROCEEDINGS{Kawai2003,
       author = {{Kawai}, N. and {Yoshida}, A. and {Matsuoka}, M. and {Shirasaki}, Y. and {Tamagawa}, T. and {Torii}, K. and {Sakamoto}, T. and {Takahashi}, D. and {Fenimore}, E. and {Galassi}, M. and {Tavenner}, T. and {Lamb}, D.~Q. and {Graziani}, C. and {Donaghy}, T. and {Vanderspek}, R. and {Yamauchi}, M. and {Takagishi}, K. and {Hatsukade}, I.},
        title = "{In-Orbit Performance of WXM (Wide-Field X-Ray Monitor)}",
    booktitle = {Gamma-Ray Burst and Afterglow Astronomy 2001: A Workshop Celebrating the First Year of the HETE Mission},
         year = 2003,
       editor = {{Ricker}, G.~R. and {Vanderspek}, R.~K.},
       series = {American Institute of Physics Conference Series},
       volume = {662},
        month = apr,
    publisher = {AIP},
      address   = {Melville, NY},
        pages = {25-32},
          doi = {10.1063/1.1579293},
       adsurl = {https://ui.adsabs.harvard.edu/abs/2003AIPC..662...25K}
}

@ARTICLE{Gehrels2004,
       author = {{Gehrels}, N. and {Chincarini}, G. and {Giommi}, P. and {Mason}, K.~O. and {Nousek}, J.~A. and {Wells}, A.~A. and {White}, N.~E. and {Barthelmy}, S.~D. and {Burrows}, D.~N. and {Cominsky}, L.~R. and {Hurley}, K.~C. and {Marshall}, F.~E. and {M{\'e}sz{\'a}ros}, P. and {Roming}, P.~W.~A. and {Angelini}, L. and {Barbier}, L.~M. and {Belloni}, T. and {Campana}, S. and {Caraveo}, P.~A. and {Chester}, M.~M. and {Citterio}, O. and {Cline}, T.~L. and {Cropper}, M.~S. and {Cummings}, J.~R. and {Dean}, A.~J. and {Feigelson}, E.~D. and {Fenimore}, E.~E. and {Frail}, D.~A. and {Fruchter}, A.~S. and {Garmire}, G.~P. and {Gendreau}, K. and {Ghisellini}, G. and {Greiner}, J. and {Hill}, J.~E. and {Hunsberger}, S.~D. and {Krimm}, H.~A. and {Kulkarni}, S.~R. and {Kumar}, P. and {Lebrun}, F. and {Lloyd-Ronning}, N.~M. and {Markwardt}, C.~B. and {Mattson}, B.~J. and {Mushotzky}, R.~F. and {Norris}, J.~P. and {Osborne}, J. and {Paczynski}, B. and {Palmer}, D.~M. and {Park}, H.-S. and {Parsons}, A.~M. and {Paul}, J. and {Rees}, M.~J. and {Reynolds}, C.~S. and {Rhoads}, J.~E. and {Sasseen}, T.~P. and {Schaefer}, B.~E. and {Short}, A.~T. and {Smale}, A.~P. and {Smith}, I.~A. and {Stella}, L. and {Tagliaferri}, G. and {Takahashi}, T. and {Tashiro}, M. and {Townsley}, L.~K. and {Tueller}, J. and {Turner}, M.~J.~L. and {Vietri}, M. and {Voges}, W. and {Ward}, M.~J. and {Willingale}, R. and {Zerbi}, F.~M. and {Zhang}, W.~W.},
        title = "{The Swift Gamma-Ray Burst Mission}",
      journal = {Ap. J.},
         year = 2004,
        month = aug,
       volume = {611},
       number = {2},
        pages = {1005-1020},
          doi = {10.1086/422091},
archivePrefix = {arXiv},
       eprint = {astro-ph/0405233},
 primaryClass = {astro-ph},
       adsurl = {https://ui.adsabs.harvard.edu/abs/2004ApJ...611.1005G}
}

@INPROCEEDINGS{Hurley2013,
       author = {{Hurley}, K. and {Mitrofanov}, I.~G. and {Golovin}, D. and {Litvak}, M.~L. and {Sanin}, A.~B. and {Boynton}, W. and {Fellows}, C. and {Harshman}, K. and {Starr}, R. and {Golenetskii}, S. and {Aptekar}, R. and {Mazets}, E. and {Pal'shin}, V. and {Frederiks}, D. and {Svinkin}, D. and {Smith}, D.~M. and {Hajdas}, W. and {von Kienlin}, A. and {Zhang}, X. and {Rau}, A. and {Yamaoka}, K. and {Takahashi}, T. and {Ohno}, M. and {Hanabata}, Y. and {Fukazawa}, Y. and {Tashiro}, M. and {Terada}, Y. and {Murakami}, T. and {Makishima}, K. and {Cline}, T. and {Barthelmy}, S. and {Cummings}, J. and {Gehrels}, N. and {Krimm}, H. and {Palmer}, D. and {Goldsten}, J. and {Del Monte}, E. and {Feroci}, M. and {Marisaldi}, M. and {Connaughton}, V. and {Briggs}, M.~S. and {Meegan}, C.},
        title = "{The Interplanetary Network}",
    booktitle = {EAS Publications Series},
         year = 2013,
       editor = {{Castro-Tirado}, A.~J. and {Gorosabel}, J. and {Park}, I.~H.},
       series = {EAS Publications Series},
       volume = {61},
        month = jul,
    publisher = {EDP},
     address   = {Les Ulis, France},
        pages = {459-464},
          doi = {10.1051/eas/1361074},
       adsurl = {https://ui.adsabs.harvard.edu/abs/2013EAS....61..459H}
}

@INPROCEEDINGS{Lund1986,
       author = {{Lund}, Niels},
        title = "{The WATCH gamma-burst detector for Eureca-I}",
    booktitle = {X-ray instrumentation in astronomy},
         year = 1986,
       editor = {{Culhane}, J.~L.},
       series = {Society of Photo-Optical Instrumentation Engineers (SPIE) Conference Series},
       volume = {597},
        month = jan,
        pages = {95-100},
          doi = {10.1117/12.966566},
       adsurl = {https://ui.adsabs.harvard.edu/abs/1986SPIE..597...95L}
}

@ARTICLE{Burns2023,
       author = {{Burns}, Eric and {Coughlin}, Michael and {Ackley}, Kendall and {Andreoni}, Igor and {Bizouard}, Marie-Anne and {Broekgaarden}, Floor and {Christensen}, Nelson L. and {D'Ammando}, Filippo and {DeLaunay}, James and {Fleischhack}, Henrike and {Frey}, Raymond and {Fryer}, Chris L. and {Goldstein}, Adam and {Grossan}, Bruce and {Hamburg}, Rachel and {Hartmann}, Dieter H. and {Ho}, Anna Y.~Q. and {Howell}, Eric J. and {Hui}, C. Michelle and {Jenks}, Leah and {Joens}, Alyson and {Lesage}, Stephen and {Levan}, Andrew J. and {Lien}, Amy and {Meli}, Athina and {Negro}, Michela and {Parsotan}, Tyler and {Roberts}, Oliver J. and {Santander}, Marcos and {Smith}, Jacob R. and {Tohuvavohu}, Aaron and {Tomsick}, John A. and {Wadiasingh}, Zorawar and {Veres}, Peter and {Villar}, Ashley V. and {Zhang}, Haocheng and {Zhu}, Sylvia J.},
        title = "{Gamma-ray Transient Network Science Analysis Group Report}",
      journal = {arXiv e-prints},
         year = 2023,
        month = aug,
          eid = {arXiv:2308.04485},
        pages = {arXiv:2308.04485},
          doi = {10.48550/arXiv.2308.04485},
archivePrefix = {arXiv},
       eprint = {2308.04485},
 primaryClass = {astro-ph.HE},
       adsurl = {https://ui.adsabs.harvard.edu/abs/2023arXiv230804485B}
}

@article{Lin2002,
  author  = "Lin, R. P. and Dennis, B. R. and Hurford, G. J. and Smith, D. M. and Zehnder, A. and Harvey, P. R. and Curtis, D. W. and Pankow, D. and Turin, P. and Bester, M. and Csillaghy, A. and Lewis, M. and Madden, N. and van Beek, H. F. and Appleby, M. and Raudorf, T. and McTiernan, J. and Ramaty, R. and Schmahl, E. and Schwartz, R. and Krucker, S. and Abiad, R. and Quinn, T. and Berg, P. and Hashii, M. and Sterling, R. and Jackson, R. and Pratt, R. and Campbell, R. D. and Malone, D. and Landis, D. and Barrington-Leigh, C. P. and Slassi-Sennou, S. and Cork, C. and Clark, D. and Amato, D. and Orwig, L. and Boyle, R. and Banks, I. S. and Shirey, K. and Tolbert, A. K. and Zarro, D. and Snow, F. and Thomsen, K. and Henneck, R. and McHedlishvili, A. and Ming, P. and Fivian, M. and Jordan, J. and Wanner, R. and Crubb, J. and Preble, J. and Matranga, M. and Benz, A. and Hudson, H. and Canfield, R. C. and Holman, G. D. and Crannell, C. and Kosugi, T. and Emslie, A. G. and Vilmer, N. and Brown, J. C. and Johns-Krull, C. and Aschwanden, M. and Metcalf, T. and Conway, A.",
  title   = "The {R}euven {R}amaty High-Energy Solar Spectroscopic Imager ({RHESSI})",
  journal = "Sol. Phys.",
  volume  = "210",
  number  = "1-2",
  pages   = "3--32",
  year    = "2002",
  doi     = "10.1023/A:1022428818870"
}

@article{Krucker2020,
  author  = "Krucker, S. and Hurford, G. J. and Grimm, O. and K{\"o}gl, S. and Gr{\"o}belbauer, H.-P. and Etesi, L. and Casadei, D. and Csillaghy, A. and Benz, A. O. and Arnold, N. G. and Molendini, F. and Orleanski, P. and Schori, D. and Xiao, H. and Kuhar, M. and Hochmuth, N. and Felix, S. and Schramka, F. and Marcin, S. and Kobler, S. and Iseli, L. and Dreier, M. and Wiehl, H. J. and Kleint, L. and Battaglia, M. and Lastufka, E. and Sathiapal, H. and Lapadula, K. and Bednarzik, M. and Birrer, G. and Stutz, St. and Wild, Ch. and Marone, F. and Skup, K. R. and Cichocki, A. and Ber, K. and Rutkowski, K. and Bujwan, W. and Juchnikowski, G. and Winkler, M. and Darmetko, M. and Michalska, M. and Seweryn, K. and Bia{\l}ek, A. and Osica, P. and Sylwester, J. and Kowalinski, M. and {\'S}cis{\l}owski, D. and Siarkowski, M. and St{\k{e}}{\'s}licki, M. and Mrozek, T. and Podg{\'o}rski, P. and Meuris, A. and Limousin, O. and Gevin, O. and Le Mer, I. and Brun, S. and Strugarek, A. and Vilmer, N. and Musset, S. and Maksimovi{\'c}, M. and F{\'a}rn{\'i}k, F. and Koz{\'a}{\v{c}}ek, Z. and Ka{\v{s}}parov{\'a}, J. and Mann, G. and {\"O}nel, H. and Warmuth, A. and Rendtel, J. and Anderson, J. and Bauer, S. and Dionies, F. and Paschke, J. and Pl{\"u}schke, D. and Woche, M. and Schuller, F. and Veronig, A. M. and Dickson, E. C. M. and Gallagher, P. T. and Maloney, S. A. and Bloomfield, D. S. and Piana, M. and Massone, A. M. and Benvenuto, F. and Massa, P. and Schwartz, R. A. and Dennis, B. R. and van Beek, H. F. and Rodr{\'i}guez-Pacheco, J. and Lin, R. P.",
  title   = "The {S}pectrometer/{T}elescope for {I}maging {X}-rays ({STIX})",
  journal = "Astron. Astrophys.",
  volume  = "642",
  eid     = "A15",
  pages   = "A15",
  year    = "2020",
  doi     = "10.1051/0004-6361/201937362"
}

@inproceedings{Makarova2002,
  author    = "Makarova, O. V. and Moldovan, N. and Tang, C.-M. and Mancini, D. C. and Divan, R. and Zyryanov, V. N. and Ryding, D. G. and Yaeger, J. and Liu, C.",
  title     = "Focused Two-Dimensional Antiscatter Grid for Mammography",
  booktitle = "Medical Imaging 2002: Physics of Medical Imaging",
  series    = "Proc. SPIE",
  volume    = "4783",
  pages     = "148--155",
  year      = "2002",
  doi       = "10.1117/12.451018"
}

@article{Makarova2008,
  author  = "Makarova, O. V. and Yang, G. and Amstutz, P. T. and Tang, C.-M.",
  title   = "Fabrication of Antiscatter Grids and Collimators for X-Ray and Gamma-Ray Imaging by Lithography and Electroforming",
  journal = "Microsyst. Technol.",
  volume  = "14",
  number  = "9",
  pages   = "1613--1619",
  year    = "2008",
  doi     = "10.1007/s00542-008-0558-7"
}

@article{Boone2002,
  author  = "Boone, J. M. and Makarova, O. V. and Zyryanov, V. N. and others",
  title   = "Development and {M}onte {C}arlo Analysis of Antiscatter Grids for Mammography",
  journal = "Technol. Cancer Res. Treat.",
  volume  = "1",
  pages   = "441--447",
  year    = "2002",
  doi     = "10.1177/153303460200100604"
}

@article{VanAudenhaege2015,
  author  = "Van Audenhaege, K. and Van Holen, R. and Vandenberghe, S. and Vanhove, C. and Metzler, S. D. and Moore, S. C.",
  title   = "Review of {SPECT} Collimator Selection, Optimization, and Fabrication for Clinical and Preclinical Imaging",
  journal = "Med. Phys.",
  volume  = "42",
  number  = "8",
  pages   = "4796--4812",
  year    = "2015",
  doi     = "10.1118/1.4927061"
}
\bibliographystyle{aasjournalv7.1}

\end{document}